# Augmenting adaptive immunity: progress and challenges in the quantitative engineering and analysis of adaptive immune receptor repertoires

Alex J. Brown[1,2], Igor Snapkov[●,3], Rahmad Akbar[●,3], Milena Pavlović[3,4], Enkelejda Miho[5,6,7], Geir K. Sandve[4] Victor Greiff[✉,3]

[1] Department of Biomedical Research, National Jewish Health, Denver, CO, USA
[2] Department of Immunology and Microbiology, University of Colorado School of Medicine, Aurora, CO, USA
[3] Department of Immunology, University of Oslo, Oslo, Norway
[4] Department of Informatics, University of Oslo, Oslo, Norway
[5] Institute of Medical Engineering and Medical Informatics, School of Life Sciences, FHNW University of Applied Sciences and Arts Northwestern Switzerland, Muttenz, Switzerland
[6] aiNET GmbH, Switzerland Innovation Park Basel Area AG, Basel, Switzerland
[7] DayOne, BaselArea.swiss, Basel, Switzerland
[●] Equal Contribution
[✉] Correspondence: victor.greiff@medisin.uio.no

## Abstract
The adaptive immune system is a natural diagnostic and therapeutic. It recognizes threats earlier than clinical symptoms manifest and neutralizes antigen with exquisite specificity. Recognition specificity and broad reactivity is enabled via adaptive B- and T-cell receptors: the immune receptor repertoire. The human immune system, however, is not omnipotent. Our natural defense system sometimes loses the battle to parasites and microbes and even turns against us in the case of cancer and (autoimmune) inflammatory disease. A long-standing dream of immunoengineers has been, therefore, to mechanistically understand how the immune system "sees", "reacts" and "remembers" (auto)antigens. Only very recently, experimental and computational methods have achieved sufficient quantitative resolution to start querying and engineering adaptive immunity with great precision. In specific, these innovations have been applied with the greatest fervency and success in immunotherapy, autoimmunity and vaccine design. The work here highlights advances, challenges and future directions of quantitative approaches which seek to advance the fundamental understanding of immunological phenomena, and reverse engineer the immune system to produce auspicious biopharmaceutical drugs and immunodiagnostics. Our review indicates that the merger of fundamental immunology, computational immunology and (digital) biotechnology minimizes black box engineering, thereby advancing both immunological knowledge and as well immunoengineering methodologies.

# Introduction

**Advancing immunology through engineering innovations**
Theodore von Kármán, the great 20th century aerospace engineer, famously stated that scientists study the world as it is, while engineers create the world that has never been[1]. Fundamental questions in science often could only be addressed *after* the development of pioneering engineering solutions. Recent examples such as the development of Taq polymerase, GFP and its derivatives, CRISPR/Cas9, high-throughput sequencing and proteomics have enabled researchers to answer countless fundamental questions in science, otherwise impossible to approach.

Advances in developing a quantitative understanding of the adaptive immune repertoire has been often brought about by biotechnology tools that have made possible performing high-throughput work. Quantitative tools have allowed us to expose enormous quantities of immune receptor sequences and their characteristics[2–15]. Emergent tools in high-throughput and synthetic biology have enabled technologies for seamless integration of gene constructs, gene circuit engineering, DNA assembly, and gene synthesis[16,17]. While many of these transformative ideas originate with the intention of solving problems pertinent to the adaptive immune repertoires, it is also true that the development of new quantitative and synthetic tools is a pervasive pursuit across many fields. The merger of synthetic and quantitative tools opens the space for not only the analysis, redesign and optimization of existing T cell receptors (TCRs) and antibodies, but also the de novo design of proteins heretofore unknown to nature. Moreover, we perceive the utility of several computational, genomics, proteomics and biotechnology tools that have not been developed with the immune repertoire in mind but are compelling candidate techniques to incorporate into the quantitative immunoengineering toolset for the purpose of the quantitative analysis and engineering of adaptive immune receptors.

Here we review and retrace works operating at the interface of immunoengineering and fundamental immunology that have solved fundamental questions in high-throughput adaptive immune repertoires and have furthered the promise of precision engineered immunotherapies and vaccine design.

**Adaptive immune receptors are natural diagnostics and therapeutics**
B and T cells together make up the adaptive immune system that detects and fights infection and disease with specificity. The power of adaptive immune cell populations is showcased by their dysregulation. Autoimmunity describes a class of diseases, such as celiac disease[18,19] or Type 1 Diabetes[20,21] in which immune cells turn against the host with disastrous consequences. The protein receptors that confer both such incredible protective and destructive capacity are B- and T-cell receptors, immune receptors. In their soluble form, secreted by plasma cells, B-cell receptors (BCRs) are called antibodies. In this review, the terms BCR and antibody are used interchangeably if not specified otherwise.

Immune repertoires are a natural diagnostics; they respond to a pathogen prior to the emergence of clinical symptoms[22]. This is the premise of most immunodiagnostic tests that use adaptive immune system dynamics as biomarkers[23]. Consequently, a long-standing goal has been to perform repertoire-based diagnostics of immune state by minimally invasive blood repertoire sequencing. The underlying assumption of repertoire-based immunodiagnostics is that a disease induces, across affected individuals, disease-associated receptor sequence (motifs) that function as disease biomarkers[24]. Recently, Emerson and colleagues were able to discriminate between cytomegalovirus (CMV) positive and negative individuals based solely on the presence of CMV-associated TCR sequences[25].

Immune receptors bind (auto)antigens with specificity leading to the neutralization through antibodies or elimination of (auto)antigens through (T cells). Upon activation, the adaptive immune system clears the eliciting antigen from the body thereby behaving as a bona fide therapeutic. In 1986, the first monoclonal antibody was FDA-approved as a treatment in kidney transplantation[26]. As of February 2019, 113 monoclonal antibodies have been approved by the US Food and Drug Administration (FDA) and/or European Medicines Agency, and hundreds more are being actively tested in clinical trials[27]. Antibodies feature among blockbuster drugs, highlighting the importance of immunotherapy for public health[27–29]. Complementarily, the first chimeric antigen receptor (CAR) T cells therapeutics have been FDA-approved in 2017 for the treatment of B-cell malignancies.

In summary, antibody and T-cell receptors are important as diagnostics and therapeutics. Therefore, enormous efforts are currently undertaken to develop tools and platforms that enable (digital) immunoengineering at a large, quantitative scale[30–32].

**Engineering the vast immune receptor sequence space requires quantitative approaches**
Directed, rule and knowledge-driven B-and T-cell engineering depends on the understanding of immune receptor biology, the structure of the antigenome and the potential interaction space that links immune receptors to antigens.

Since the first publications on high-throughput immune repertoire sequencing[33–35], the diversity of immune receptor repertoires has been described in extensive detail by sophisticated experimental and computational approaches as also reviewed by others[2,3,10,36–40]. For example, the number of immune receptors stored in the iReceptor database, a searchable public portal for immune repertoire studies, recently exceeded 1 billion[7]. On the single individual level, a new frontier in sequencing depth has been recently established with $10^7$ sequencing reads per individual closing in on the number of lymphocytes in the blood ($10^8$)[41]. Similarly, antigen landscapes have been investigated using both mass spectrometry and display technologies (e.g., phage display, peptide and protein arrays)[42,43]. These investigations have led to large epitope databases such as the Immune Epitope Database (IEDB) that include more than validated 130,000 peptide epitopes[43]. Thus, the charted landscape of immune receptor sequence and epitope diversity is constantly expanding.

Despite ever growing data resources, the ability to link immune receptor and antigen sequence, at high-throughput, has only recently begun to be investigated. The complexity of immune receptor diversity, the antigenome and their interaction space, requires experimental and computational analytical approaches to identify and detect engineerable interaction patterns and rules. In this review, we define quantitative immunoengineering as the set of high-throughput, high-precision genomic, proteomic, computational and biotechnology tools that have jointly created an unprecedented opportunity to repair, (re-)create and improve human adaptive immunity.

## Current approaches for immune repertoire analysis and immunoengineering

**Computational immunology and immunoinformatics of adaptive immunity**
Computational immunology is of crucial importance for understanding the complexity of adaptive immunity[11]. De novo immunoengineering of adaptive immunity requires (i) knowledge and replication of germline gene recombination, (ii) understanding of disease and antigen-specific patterns in immune repertoires, (iii) analysis of B-and T-cell population dynamics, and (iv) modeling of immune receptor 3D structure. In these four areas, remarkable computational progress has been made over the past years. We believe that the computational progress described here is needed to analyze ultra-large immune receptor ($>10^7$ clonal sequences) datasets with greater efficiency[44].

**Immune repertoire *in silico* simulation and diversity**
The rules of V(D)J recombination shaping immune repertoire structure have been under intense investigation since the seminal paper by Mora and Walczak in 2010[45]. Throughout a series of papers, the groups of Walczak and Mora have showed that V(D)J recombination in B and T cells may be modeled by Hidden Markov models[46] as well as Bayesian probabilistic methods[47]. These models of VDJ recombination, trained by repertoire sequencing data, have been applied to the investigation of: (i) potential repertoire diversity and estimation of receptor sequence generation probabilities[45,48,49], (ii) repertoire selection[50–52], (iii) the phenomenon of naïve public clones and the identification of antigen-specific public clones[47,53,54], (iv) and for the generation of nature-like immune repertoires[47,55].

Mora, Walczak and other groups found that the potential mouse and human naïve B- and T-cell diversity is at least $10^{14}$ [48,49,56,57], respectively. Despite the immense repertoire diversity, we know of many instances of individuals sharing similar or identical public immune receptors with each other[8,9,25,56,58–61]. Functionally, these patterns are likely to be public markers of evolutionary memory, immunological memory and human leukocyte antigen (HLA), with the most highly correlated clusters strongly linked to common viral pathogens[62,63]. The authors showed that naïve public clones are well explained by power-law distributed VDJ recombination statistics. The computational tools IGoR[64] (nucleotide level) and Olga (amino acid level)[55] implement methods for estimation of VDJ recombination scenarios as well as generation of in immune receptor sequences given an germline gene sets and inferred recombination statistics. Using IGgoR, Briney and colleagues have showed that inferred VDJ recombination models differ across individuals. This suggests that the practice of using one general of VDJ recombination across all individuals might yield less accurate results[8]. This finding is in line with the emergence of personalized germline gene repertoires through the existence of immunoglobulin germline gene polymorphisms[65–68]. An increasing number of antibody polymorphisms[67,69] warrants also the question of how repertoire variation on the germline gene level translates to the creation of antigen-specific immune antibody repertoires[65].

Multiple reports have provided evidence for a power-law distribution of clonal diversity[45,70–72]. It has been previously suggested that power-law distributions arise in the presence of hidden unobserved variables, such as that of pathogens[73]. Complementarily, Desponds and colleagues showed that power-law distribution of lymphocyte clones can arise solely due to fitness fluctuations. In agreement with Schwab and colleagues[73], Desponds and colleagues found that power-law distributions can result entirely from an exogenous environment with which the adaptive immune system interacts.

Given the power-law nature of clonal frequency distributions (*p*), one can assume that they contain immunological information. In order to capture this information, clonal frequency distributions have to be converted to a common reference framework that is clone-independent. Clone-independence is important since there is little overlap across individuals (despite there being considerable numbers of public clones[8,9,56]). Such clonal independence was achieved with the help of prior work in mathematical ecology. Specifically, calculating the exponential of the Rényi entropy, called Hill-Diversity profiles, $((\sum_{i=1}^{N} p_i^q)^{\frac{1}{1-q}})$, across an array of *q*-values yields a lower-dimensional vector that was shown to capture a large extent of the information contained in clonal frequency distributions[70]. Hill-diversity profiles capture the state of clonal expansion of any given immune repertoire and could discriminate mouse immunization, health, transplantation and cancer statuses, as well as lymphocyte populations[56,70,74].

While diversity profiles capture the information contained in the clonal frequency distribution, they do not capture sequence-dependent information. Two clonal repertoires may have identical clonal frequency distributions but may be composed of entirely different clonal sequences. Sequence-dependent information is captured by the sequence similarity landscape of immune repertoires[53,63,75–77]. Specifically, the similarity landscape of a repertoire may be determined by calculating the sequence distance matrix [Levenshtein distance] of an immune repertoire. Using this distance matrix, a network can be drawn wherein each clonal sequence is a node and each edge indicates an editdistance (number of changes to convert one nucleotide or amino acid sequence into another sequence) between clones. Such networks may be drawn for both B cells[76,77] and T cells[63]. If calculated across several distance cut-offs, these networks may be used to compare distinct layers of repertoire similarity, which has offered unprecedented insight into the architecture of repertoire similarity[77]. For example, Miho and colleagues found that these layers are correlated, suggesting that similarity in repertoires is organized according to underlying but, as of yet, unidentified rules. Another striking observation in immune repertoires is that public clones are biased to higher frequency and seem to represent network hubs[63,77] surrounded by private clones[77]. Finally, Priel and colleagues recently published a method for following network architecture of T-cell receptor repertoires over time as well as analyzing it via machine learning[78].

While inspecting the similarity relation landscape of a repertoire provides crucial insight into repertoire biology, computational demands for investigating repertoire similarity may become prohibitive for larger repertoires. Indeed, networks of repertoire sizes larger than 100,000 clones require either cloud services or parallelized computational frameworks on large-scale computing clusters as recently developed by Miho and colleagues[77]. By calculating networks of millions of clones, Miho and colleagues showed that the practice of inspecting a portion of a repertoire (subsampling) may be acceptable, depending on the number of public clones in the subsampled portion. In fact, random samples of up to 50% of the original repertoire-maintained network properties similar to the originating repertoire. In this regard, network analysis and diversity profiles show similar robustness to undersampling[70,77].

Clonal frequency and the sequence similarity landscape contain complementary immunological information. In recent work, Arora and colleagues as well as Strauli and colleagues published mathematical frameworks allowing the merging of sequence and frequency information [79,80], however it remains undetermined how immunological information is distributed among frequency and sequence-based repertoire components[81,82].

In summary, we can now generate *in silico* immune repertoires and quantify repertoire diversity thanks to advances in probabilistic modeling, mathematical ecology and network theory.

**B-and T-cell pattern mining using machine and deep learning**
Immune receptor specificity is one of the hallmark features of adaptive immunity and encoded into the sequence of each immune receptor in motifs. An underlying assumption of pattern mining is that receptors that are specific for a given antigen, share similarity of certain motifs.

Davis, Thomas and Laukens' groups published seminal papers regarding T-cell receptors showing that prediction of TCR antigen specificity can be performed based on the TCR sequence alone with high accuracy (≈80%)[24,83–85]. The authors used distance-based classifiers (TCRDist[24] and GLIPH[83]) and other sequence features such as complementarity-determining region 3 (CDR3) length and CDR3 physicochemical properties[84]. Recently, Jokinen and colleagues published a method based on Gaussian processes (called TCRGP) that, in contrast to TCRDist[24], does not assign a priori weights to CDR regions when predicting antigen specificity on the entire VDJ region[86] improving on the performance of TCRDist by 6% percentage points (86% prediction accuracy). Distance-based clustering approaches for epitope-specific clustering of immune receptors were formally benchmarked in reports from Meysman as well as Thakkar and colleagues that investigated different distance metrics and alignments for TCR-based clustering. These studies indicate that, at least for TCR-based analysis, distance-based measures are well suited for finding commonalities among epitope-specific TCRs, disease-associated TCRs and TCR across individuals (such as twins). However, simple distance-based measures render differentiation between short and long-range sequence interaction unfeasible. Specifically, outlier sequences that are very dissimilar to the main sequence cluster but still bind to the same epitope are not captured by these distance-based classifiers. These outlier sequences remain a challenge for future research – both for supervised and unsupervised approaches[87]. As of yet, it remains unclear to what extent short and long-range interactions govern major histocompatibility complex (MHC)-TCR-antigen interaction. Recent data by Ostmeyer indicate that interaction of TCR and peptide is well-captured by short and continuous sequence stretches[82].

For antibodies, there exist fewer approaches for sequence-based prediction of antigen specificity. Previously, two approaches have been published that involve deep learning of paratope and paratope-epitope coupled prediction of antibody-antigen binding, respectively[88,89]. Using the structural information from antibody-antigen complexes, Liberis and colleagues built a deep learning classifier based on convolutional and recurrent neural networks that incorporated both amino acid sequence and physico-chemical property information. Using their deep learning classifier, they were able to predict paratope residues with an F1-score [a measure of prediction accuracy] of 0.69. Importantly the deep learning model also inferred well the binding frequency of each amino acid in a paratope. In a follow-up manuscript, the authors substituted the recurrent layer with convolutional layers specifically designed to cover longer range sequence interaction as well as a cross-modal-attention layer of antibody over the antigen residue features[88]. Interestingly, by incorporating antigen information into the paratope predictor, prediction results were only improved slightly. This may be due to the limited number of antibody-antigen complexes currently available (≈800)[90].

Greiff and colleagues used sequence motifs to predict sequence publicity of BCR and TCR clones[91]. Leveraging a gapped-k-mer SVM approach[92], they found that there exist motifs that predict both in mouse and human, immune receptor publicity with 80% prediction accuracy. The classifier was tested to perform well across datasets and sequencing library preparation methods. This classifier is complementary to the public/private TCR clone classifier "PUBLIC", which accurately predicts shared TCR clones between two people or an arbitrarily large number of individuals[54]. Specifically, based on T-cell immune receptor generation probabilities, and extrinsic factors such as cohort size, sequencing depth, and a definition of what constitutes as a public receptor, the Walczak group suggests that the degree of publicity of a given TCR clone is largely dependent on the probability of its generation through V(D)J recombination.

While over the last few years there has been considerable interest in the classification of antigen or epitope-specific sequences, there has been relatively less conceptual progress on the classification problem of the immune status based on entire repertoires (per individual samples). Briefly, there is a conceptual difference in machine learning between sequence-based (prediction of antigen specificity) and repertoire-based classification (prediction of immune status). While in the sequence-based case class labels are assigned in a binary or multi-class fashion to a *single-sequence* (for example: sequence 1→ class 1 – binding to HIV, sequence 2 → class 2– binding to Tetanus), labels in the repertoire classification case are given to a *set of* sequences. The assumption is that within a repertoire there exist sequences or subsequences (single or sets of sequence motifs) that are overrepresented in one class compared to all other classes. The particular challenge in this case is that the signal-to-noise ratio in immune repertoires is unknown but likely very low (although, for example, in the case of CMV, nearly 10% of T-cells are CMV-specific[93]). Specifically, the disease signal in the form of sequence (motifs) likely represents a high-dimensional mixture of different motifs. Machine learning approaches to disentangle and recover these high-dimensional signals are as of yet missing[94]. Nevertheless, progress has been made in terms of repertoire-based diagnostics in mouse and humans with both model antigens and virus infections leveraging both sequence motifs and entire sequences[25,82,95–98]. Machine learning approaches for repertoire-based diagnostics range from Bayesian classifiers, support vector machine and very recently multiple instance and deep learning. While all of these approaches show considerable prediction accuracy, the identification of disease-status driving sequence or sequence motifs remains difficult. Emerson and colleagues solved this problem by exclusively working on public clones and consequently identifying clones that were associated with CMV status using Fisher's exact test[25]. However, there might also be considerable disease-associated information in the private portion of each repertoire. It remains a general and important problem how immune-relevant information, that is seemingly variable within classes and thus private, may be leveraged for disease-state classification[99]. In addition to disease or antigen-specific information spread among public and private parts of the repertoire, several studies have shown an HLA-effect on repertoire structure with public clones or germline genes being specifically linked to certain HLA-types[25,58,100]. It remains unclear to what extent HLA-induced repertoire structures would bias disease-specific repertoire classification and to what extent repertoires ensembles are characteristic of a certain HLA type[101,102].

More generally, given that pathogen diversity drives human MHC evolution[103] and the personalized nature of epitope repertoires[104], certain pathogen groups influence MHC evolution more than others depending on the environment. Thus, as adaptive immune repertoire analysis moves towards more clinically focused questions and larger cohorts, evolutionary and epidemiological approaches will be needed to appropriately take covariates such as environment, age[105–108] and genetic confounders[58] into consideration when performing privacy-preserving and ethics-conform disease-specific immune receptor-based classification[109]. Ostermeyer and colleagues found that their classifier generalized well across individuals with presumably different HLA backgrounds[82]. The authors hypothesized this because of two reasons: (1) TCR-MHC interaction occurs primarily via CDR1 and CDR2 whereas peptide contacts are primarily via CDR3 and (2) their study design included HLA-matched controls[82].

Of note, while HLA-restriction of *T-cell* repertoires has garnered increased attention, investigations as to a potential HLA-restriction of antigen-specific *B-cell* repertoires has not been explored to our knowledge. B-cell activation and germinal center reaction are, in part, T-cell dependent, which may have an impact on which B cells receive appropriate signals to participate in the humoral immune response.

**Mathematical modeling of immune receptor recognition**
Manipulation of adaptive immunity by either *in vitro* immunoengineering of immune receptor repertoires[110–112] or *in vivo* (no reports yet to our knowledge) depends on the underlying recognition dynamics of immune cells. One of the earliest theoretical studies on immune receptor recognition was performed by Perelson and collaborators summarized comprehensively by Perelson and Weisbuch[113]. Perelson and Weisbuch worked on the mathematical modeling of the immune system from the viewpoint of how such a complex system can achieve recognition and memory of a nearly limitless number of antigens by at the same time being restricted by (i) immune cell number, (ii) B and T-cell clone size and (iii) immunogenetics (germline gene repertoire). So far, the maximum number of antigens to which the human immune system can build a memory response remains unclear[114]. Specifically, one of the main questions of Perelson and Weisbuch is the problem of repertoire completeness – where completeness describes the extent to which immune repertoires have the potential to recognize any given antigen. Repertoire completeness is an old problem that has received relatively little attention as of late, despite its importance for immunoengineering. Indeed, prior to modifying the binding landscape of immune repertoires via *in vitro* or *in vivo* engineering, detailed knowledge of the a priori immune receptor binding landscape is needed.

To mathematically simulate binding of antibodies to antigen, bit strings (string models) are often employed to represent antibodies as well as antigens[113,115–117]. Antibody and antigen sequences are only composed of two "amino acids", 0 and 1 (binary models). The patterns of the bits represent the shapes of molecules and determine their ability to bind with other molecules[115]. In the bit string methodology conceived by Farmer and colleagues, molecular binding takes place when the bit strings of antibody and antigen "match" each other. A match occurs when the antigen and antibody have complementary binary patterns[116]. Interestingly, in their work, Farmer and colleagues attributed the ability of the immune system to learn, retain memory, and recognize patterns (antigens) to its capacity to employ genetic operators such as gene rearrangement and mutations without a priori programming. They developed a (dynamical) model that simulates the behavior of an immune system *in silico*. The model shares many commonalities with a general purpose machine learning/artificial intelligence algorithm that was being introduced in the same year by J.H. Holland called the classifier system[118]. The similarities not only include the absence of a priori programing to achieve the corresponding final goals (immune system: antigen recognition; general learning: class discrimination) but also extend component-wise. For example, in the classifier system the *classifier, condition,* and *action* components are likened to *antibody type, epitope, and paratope* components of the immune system (for detailed component-wise comparison, see Table 1 in Farmer et al.[115]). In a recent article, Efroni and Cohen further expand on the idea of the immune system "classifying" and "computing" the state of the body in a fashion similar to machine learning[119].

Bit string approaches were also explored *in vitro*: for example, Fellouse and colleagues[120] obtained functional antibodies from a library of antigen-binding sites generated by a binary code restricted to tyrosine and serine. An antibody raised against human vascular endothelial growth factor recognized the antigen with high affinity and specificity in cell-based assays.

A challenge for shape space (see Focus Box 2) and binary models is their discretization of the affinity space[121]. An overview of alternative models was recently summarized by Robert and colleagues[121]. For example, string model approaches aim to simulate antibody antigen recognition in 3D-space[121] now enable the modeling of cross-reactivity. Specifically, structural representations of amino acid sequences on a 3D grid have been proposed using an experimental interaction matrix between each pair of amino acids[122]. The respective best-folding structure of antibody and antigen is computed for both proteins, possible binding interfaces are computed, and the best affinity is recorded.

Recently string models have been applied to research questions ranging from B-cell epitope prediction to understanding multi-epitope recognition. For example, Greiff and colleagues used string models to simulate the binding of arbitrarily large antibody mixtures to peptide libraries and showed that both antibody repertoire diversity and antibody concentration within a repertoire may be crucial to humoral immune recognition[123]. Furthermore, Wang and colleagues recently used string models to model selection forces during germinal center affinity maturation. Specifically, they used such *in silico* models to gain mechanistic insights into the processes of affinity maturation that are induced by multiple antigen variants, and to compare the predicted relative efficacy of different immunization schemes in inducing cross-reactive broadly neutralizing antibodies. They found that induction of cross-reactive antibodies often occurs with low probability because conflicting selection forces, imposed by different antigen variants, can "frustrate" affinity maturation. There exist competing theories about frustration which are rigorously summarized by Robert and colleagues[121]. In another example of string modeling, Luo and colleagues generated new hypotheses regarding the difficulty of predicting the emergence of broadly neutralizing antibodies. By modeling antibodies as single strings and viruses as double strings (simulating one variable and one conserved epitope on the HIV envelope protein), they found that broadly neutralizing antibodies could in fact emerge earlier and be less mutated, but that they may be prevented from doing so as a result of competitive exclusion by the autologous antibody response. If less mutated broadly neutralizing antibodies exist, they posit, it may be possible to elicit them with a vaccine containing a mixture of diverse virus strains[124]. In summary, even seemingly simple string model scan provide insight into engineering optimal therapeutic antibodies and vaccines by modeling the paratope-epitope interface.

**Computational modeling of immune receptor 3D structure**
The majority of immune receptor sequence studies are performed on the linear (2D) V(D)J sequence, or a sub region thereof (mostly CDR3). However, immune receptor antigen interaction is inherently 3-dimensional. Consequently, residues that are distant in the linear sequence space may be very close to one another in the 3D space due to folding. Computational prediction of immune receptor structure is important in view of the unresolved technical and resource-imposed challenges related to experimental immune receptor structure determination. Rational therapeutics and vaccine design depend on 3D-modeling of immune receptor antigen binding pairs. In this review, we focus on (3D) *antibody* structure and *antibody-antigen* interaction. For *TCR*-based 3D modeling and engineering, we point the reader to recent works on the subject[125–131].

In the 3D-space, the CDRs form loops. These loops can be categorized into canonical conformations depending on the 3D-length. However, the longer the CDR the harder it becomes to predict spatial

conformation – especially that of the CDR3[132]. Typically, the CDR-H3 loop does not adopt canonical conformations and must be modeled *de novo* for maximum accuracy. In addition, the H3 loop lies at the interface of the two variable domains of the heavy ($V_H$) and light chains ($V_L$) and can interact with residues on either chain. To account for these interactions, as well as the overall geometry of the paratope, the $V_L$-$V_H$ orientation is usually optimized during H3 modeling. Accurately modeling CDR H3 and the $V_L$-$V_H$ orientation are typically the most challenging and critical aspects of antibody structure prediction[133,134].

The Gray lab has substantially contributed to computational 3D antibody structure calculations by co-developing antibody structure determination within the Rosetta software framework[134,135]. Briefly, Rosetta antibody structure modeling involves identification of the most homologous template structures for heavy and light framework regions and each CDR loop. Subsequently, the most homologous templates are assembled into a side-chain optimized model. For a high-resolution model, additional modeling of the hypervariable CDR-H3 loop is performed by also relieving steric constraints optimizing the CDR backbone torsion angles and perturbing the relative orientation of the heavy and light chains. Based on the Rosetta Framework, the Gray lab has also developed an antibody-antigen docking software (SnugDock[134,136]). It takes as input antibody-antigen structures. Specifically, SnugDock simulates the induced-fit mechanism through simultaneous optimization of several degrees of freedom. It performs rigid-body docking of the multibody ($V_H$-$V_L$)-Ag complex, as well as remodeling of the CDR-H2 and -H3 loops.

While the RosettaAntibody software may use up substantial computing hours to perform antibody structural modeling (≈1000 CPUh/antibody) due to *ab initio* modeling, the Charlotte Dean lab has developed an alternative antibody structural modeling approach based on database-search driven homology modeling, which enables CPUh-saving modeling of large numbers of antibody sequences[137] relying on the structural conservation of antibody shapes. Specifically, Kawczyk and colleagues structurally mapped 35 million antibody sequences from 600 individuals. They have developed a structural annotation of antibodies (SAAB) algorithm to bridge the sequence-structure gap in antibody repertoire analysis. Given a fasta file of antibody sequences (may be unpaired), SAAB maps the full sequences, frameworks, and CDRs to the high-quality antibody structures currently available in the Protein Data Bank (PDB). The Dean lab associated a majority of frameworks and CDR sequences to an existing antibody structure, thereby recapitulating on a large scale that the observed antibody sequence space appears to employ only a conservative set of structural shapes. The SAAB algorithm is based on the methods ABodyBuilder (antibody modeling pipeline based in part on FREAD][138] and FREAD protocols –loop structure prediction using a database search algorithm)[139] that produce models of the variable regions and protein loops.

In summary, *in silico* prediction of antibody structure is possible with reasonable accuracy and speed, at least for antibodies with moderate CDR3 length.

**Computational modeling of antibody-epitope interaction**
It is generally thought that the CDR3 is the most important region for antigen binding[140,141]. However, at least for antibodies, it has been found that the framework regions can contribute up to 20% to antigen binding thereby expanding our view of what constitutes the antibody antigen binding site[142]. Furthermore, not all the residues within the CDRs bind the antigen. Only up to 5 of these amino acids dominate in terms of binding energy. In both epitope and paratope, substitutions both in and away from the binding site can change the spatial conformation of the binding region and affect the binding reaction[143–146].

In general, antibody CDRs have a much greater frequency of tyrosine and tryptophan residues than usually found on the surface of protein molecules[143,147]. These aromatic side chains can make large rotations with little entropic cost, and they contribute significantly to the binding energy[148]. Furthermore, crystallographic studies showed that binding involved a certain amount of induced fit[149]. Upon binding, residues are displaced by several angstrom[150]. In fact, two molecules that have nearly identical structures on the basis of crystallography may not interact comparably with a given receptor because of differences in molecular dynamics[151]: the crystallographic structure of an antibody-antigen complex captures merely one point in time. The contributions of the time dimension should therefore, ideally, be taken into account for a characterization of bimolecular interactions[152,153]. Hence, Greenspan proposed a richer epitope description by taking into account (i) the spatial coordinates of the contact atoms, (ii) the dynamics in time of the atoms involved in contact with the paratope, (iii) the relative energetic contributions of atoms or amino acids to the interaction or to the discrimination between cognate epitopes and other epitopes, and (iv) the context in which the binding takes place[152,154].

While specificity has been primarily discussed in the literature in the context of monoclonal antibodies, *monoclonal* specificity does not per se explain *humoral* specificity. Indeed, despite antibody polyspecificity (cross-reactivity), the population of serum antibodies as a whole shows a high degree of specificity towards the eliciting antigen[155]. Serum specificity provides the very basis for the clearing of pathogenic agents from the body. Talmage suggested that "in a mixture of a large number of different globulin molecules, the dominant reactivity will be that common to the largest number of molecules present"[156]. Serum specificity may therefore be regarded as an ensemble phenomenon of serum antibodies [123,155,157,158].

The ability to predict B-cell epitopes for a given protein precedes digitized vaccine design[159] and diagnostics. Epitope prediction is often done by immunizing the host with the antigen followed by profiling the resulting serum-antibody response with a wide array of methods. Some of these are shortly outlined in this section. Although it is believed that the majority (>90%) of B-cell epitopes are discontinuous epitopes[144], the experimental determination of epitopes has focused primarily on the identification of continuous B-cell epitopes[160]. Apart from X-ray crystallography, which represents a structural approach to epitope mapping[90,161,162], important experimental techniques to map epitopes (and receptor cross-reactivity) are phage-display libraries[163], peptide arrays[164–167] and site-directed mutagenesis[168].

Several computational methods for prediction of continuous B-cell epitopes have been published in recent years. Propensity scale methods assign a propensity score to each amino acid that measures the tendency of an amino acid to be part of a B-cell epitope (as compared to the background)[169–174]. However, Blythe and Flower have assessed 484 amino acid propensity scales to examine the correlation between propensity scale-based profiles and the location of linear B-cell epitopes in a dataset of 50 proteins. Their study showed that even the best combinations of amino acid propensities yielded B-cell epitope predictions that were only marginally better than random[175,176]. Due to the poor results yielded by propensity scales alone, several authors have explored methods for improving the predictive performance of propensity scale methods by combining them with Hidden Markov models or support vector machines[177,178]. However, the combination of scales with several machine learning algorithms showed little improvement over single scale-based methods[179,180].

Given that immunogenicity of proteins is poorly understood, it remains an open question whether B-cell epitopes could be deciphered as intrinsic features of proteins after all[176,181]. One possible explanation for the failure of B-cell epitope prediction methods based on amino acid characteristics is that the amino acid differences between epitopes and other residues are not substantial. In line with this argument, several analyses[145,182–185] have shown that the amino-acid composition of epitopes is essentially indistinguishable from other surface-exposed non-epitopic residues. This lack of intrinsic properties to clearly differentiate between epitopic and non-epitopic residues and the fact that most of the antigen surface may become a part of an epitope under some circumstances[179,186] suggest that epitopes depend, on the antibody that recognizes them, as suggested above . Indeed, T-cell epitope prediction methods are MHC- and thus context-dependent[187].

Recently, B-cell epitope prediction methods have been proposed that take into account also the antibody sequence. These methods are motivated in part by: (i) the success of partner-specific protein-protein interface predictors[188,189] as well as allele-specific MHC binding site predictors[190,191]; and (ii) the observation that virtually any surface accessible region of an antigen can become the target of some antibody and elicit an immune response[192,193] hence requiring focus of epitope predicting for a given specific antibody. Antibody-specific B-cell epitope prediction methods take into account the binding antibody sequence or structure in order to predict conformational B-cell epitopes in a query antigen sequence of known structure[194–198]. Very recently, Jespersen and colleagues identified several geometric and physicochemical features that are correlated in interacting paratopes and epitopes, used them to develop a Monte Carlo algorithm in order to generate putative epitopes-paratope pairs, and train a machine-learning model to score them. The authors showed that, by including the structural and physicochemical properties of the paratope (e.g.,: hydrophobicity, size, Zernicke moments[199]), they could improve the prediction of the target for a given BCR.

Very recently, the Wardemann and Julien labs have uncovered that affinity maturation does not only increase diversity and affinity[200] but also homotypic interaction between repeat-bound monoclonal antibodies[201]. Somewhat relatedly, the Högberg lab has investigated isotype-specific differences in binding to nanopatterned antigen showing that antibody affinity changes with antigen-to-antigen-distance[202]. Thus, the complexity of antibody-antigen interaction is even higher than previously thought.

Taken together, modeling of immune receptor-antigen interaction still requires extensive investigation, most importantly in the areas of large-scale experimental generation of receptor-antigen pairs as well as machine learning methods that enable to define the rules of receptor-antigen interaction.

**Genomic sequencing of immune repertoires**
High-throughput immune receptor repertoire sequencing has so far mainly revolved around single-chain characterization (either α or β for T cells and predominantly $V_H$ for B cells) due to the fact that standard «bulk» sequencing approaches fail to capture natural pairing of α/β and heavy/light chains because they are encoded by separate mRNA transcripts. Chromosomal separation of B-and T-cell chain loci has made single-cell immune receptor sequencing an extremely challenging and laborious task. Nevertheless, the information on cognate pairing is essential for the structural modeling and further immune receptor engineering as the entire receptor complex determines stability, conformation, half-life, and specificity[203–205].

Until recently, the wide use of single-cell sequencing techniques has been impeded by several factors, such as low yield, low throughput and high costs. Introduction of microfluidic devices allowing to separately analyze hundreds of thousands of cells significantly improved analytical capabilities[206]. There is still a major difference in throughput between microfluidic platforms varying from 2–3 to millions of isolated single cells per run (Table 1). However, rapid evolution of the technology, along with decreasing costs, may soon provide the possibility for routine single-cell sequencing with a magnitude comparable to conventional bulk sequencing (millions of input cells).

After single-cell capture and subsequent cell lysis, mainly two strategies are currently used to retrieve full-length paired receptors. (i) Receptor chains may be paired by means of overlap-extension PCR within droplets as it was demonstrated by DeKosky and colleagues for BCRs and by Turchaninova and colleagues for TCRs[207,208]. The main disadvantage of such approaches is Illumina read-length limit (2x300 bp) that has prevented so far, the characterization of *full-length* paired sequences. (ii) An approach that might overcome this issue relies on the separate labeling of mRNA transcripts by a set of unique molecular identifiers (UMIs). These UMIs might be directly co-encapsulated along with cells[209], delivered into droplets being attached to a bead surface as it is performed in protocols by 10xGenomics, or appended to transcripts through a sequence of labeling steps[210]. Although UMI usage is associated with a number of technical challenges (including the necessity of precise single-molecule dilution and complex design of several sequential in-droplet PCRs) and relatively high costs[4], this technique has dramatically improved error correction of single-cell sequencing, thereby improving data reliability and confidence[211–213].

The understanding of antigen-specific immune repertoires remains limited due to the current technological challenges to link immune receptor and cognate antigen at high-throughput. While single-cell analysis has gained momentum in the last few years and many sequencing technologies that have been developed to advance our understanding of antigen-specific immune repertoires, not one method has emerged as a breakthrough technology with wide adoption. There is considerable variability in the cost, ease of implementation, throughput, and depth of information that can be generated from these single-cell immune repertoire sequencing and selection methods. To illustrate this point, in Table 1 we summarize the majority of the most recently published single-cell sequencing protocols currently in use, and their key differences.

**Identifying candidate TCRs or antibodies via high-throughput library screens**
In the domain of antibodies, major advances have come out of in vitro techniques to express and select epitopes by phage and yeast display. Notably, Andrew Bradbury's work on developing a recombination system in singly infected bacteria expressing Cre recombinase provided a simplified means of generating functionally rearranged $V_H/V_L$ libraries with diversities of $>10^{11}$ [214]. Generation and screening of these combinatorial display libraries has rapidly accelerated the process of antibody development. As a result, recombinant antibody-based products have entered the biopharmaceutical market[28,215,216]. Through the integration of paired antibody sequencing and yeast display libraries, new high throughput pipelines can now be used to interrogate millions of natively paired antibodies in humans[217]. Germline targeting of condition-specific antibodies, paired with high-throughput library screens, has also shown promise in detecting broadly neutralizing antibodies in HIV and is further described in Section *Setting targets on public and private immune receptors* of this review.

In the context of TCRs, efforts to identify and interrogate TCR targets and cross recognition have been primarily accomplished through barcoded peptide-major histocompatibility complex (pMHC) multimers[218–221] and yeast display libraries[125,222,223]. Recently Gee et al. crucially demonstrated that pHLA-A∗02:01 yeast display libraries (in the order of ~$10^8$ unique peptides) could be used to in a non-biased manner screens for high-affinity orphan TCRs derived from tumor infiltrating lymphocytes expressed on human colorectal adenocarcinoma[223]. Zhang and colleagues have also developed TetTCR-seq to link TCR sequences to their cognate antigens in single cells at high throughput by exploiting DNA-barcoded pMHC tetramers[224]. With this technology, the authors were able to readily investigate cross-reactivity with neo-antigens and to rapidly isolate neoantigen-specific TCRs with no cross-reactivity to the wild-type antigen. TetTCR-seq would also integrate with single-cell transcriptomics and proteomics to investigate links between single T-cell phenotype, TCR sequence and pMHC-binding. Relatedly, Bentzen and colleagues developed two approaches that enables in-depth characterization of TCR recognition patterns that govern pMHC interaction[218,219]. Specifically, they used DNA barcode-labeled MHC multimers, which allows the simultaneous study of the interaction of one clonal TCR with multiple related pMHC epitopes. By measuring TCR affinity to bound pMHC, this workflow also shows promise in predicting and validating cross-reactivity of TCRs[219]. Assessing cross-reactivity is an essential step in validating any candidate TCR in clinical development. However, directly identifying cross-reactivity to the human proteome, at the time of candidate TCR identification, is an important technology advance. Wide implementation of this practice could decrease the time needed to validate new TCRs being considered for use clinically. A major limitation to the yeast display libraries and barcoded pMHC approaches, however, is the requirement of either soluble TCR or soluble pMHC, which does not directly mimic TCR-pMHC in a physiological setting.

Two approaches capable of deciphering TCR specificity using cell-cell binding interaction, have recently been developed by the Kopf and Baltimore groups[225,226]. Both approaches attempt to identify TCR specificity utilizing a chimeric HLA extracellular domain fused to TCR downstream signaling elements and signaling via CD3ζ and an NFAT reporter. Key differences between these two approaches is that Kopf's

MHC–TCR chimeric receptors identify orphan CD4$^+$ T cell epitopes via extracellular domains of MHC-II[225], and Baltimore's platform targets CD8$^+$ T cells epitopes through MHC Class I extracellular domains[226]. While the latter has the disadvantage of decreased sensitivity (detection sensitivity of 1:10$^4$ vs. 1:10$^6$), the approach can be uniquely performed in a single round of selection.

Collectively these approaches highlight the challenge of identifying candidate TCRs or antibodies, and the tradeoffs in terms of detection sensitivity, library, diversity and ability to accurately replicate physiological binding interactions.

**Proteomic sequencing and serological profiling of antibody repertoires**
Remarkably, despite major advances in immunology research and sequencing strategies, comprehensive knowledge of the circulating antibody repertoire including specificity, function, and clonality remains scarce. Indeed there is no consensus on the quantity of distinct antibody types in the blood with numbers estimated from 10$^3$ to 10$^6$ [227,228]. This knowledge gap significantly hinders extensive understanding of serological humoral immunity, such as for example vaccine-induced antibody dynamics.

The immense antibody diversity, the little overlap across individuals, combined with high structural immunoglobulin similarity, renders the characterization of the serum antibody repertoire challenging. As outlined by Wine and colleagues, methods to study the circulating antibody repertoire may be divided into two major groups: phenotyping (e.g. ELISA, serum electrophoresis, etc.) and deciphering (LC-MS/MS )[228]. While the techniques from the former set of methods dominated immunology for decades, they are unable to grasp the sequence diversity of serum antibody repertoires. Availability of LC-MS/MS (mass spectrometry) technology with its high throughput (and previously unachievable specificity and sensitivity) enabled characterization of circulating immunoglobulins with resolution superior to any other analytical methods. Specifically, in 2012, Cheung and colleagues published the first paper describing direct identification of antibodies (CDR-H3 peptides) from serum using high-throughput B-cell receptor sequencing data as a reference[229]. Recently, Lee and co-authors described a novel class of non-neutralizing antibodies abundantly presented in the circulation of individuals immunized with influenza vaccine[230]. They also found in a human donor, of which the influenza serum antibody response was followed for several years, that 24 persistent antibody clonotypes accounted for ≈70% of the anti-H1N1 A/California/7/2009 repertoire indicating the astounding stability of the serum antibody repertoire[231].

It is worth mentioning that in order to exploit its full analytic potential, LC-MS/MS technology requires robust reference datasets generated by high-throughput sequencing, although the number of tools for *de novo* sequences identification is increasing[232]. Moreover, overall high costs and technical complexity along with additional infrastructure requirements currently prevent broad application of antibody serum LC-MS/MS.

While mass-spectrometry-driven antibody proteomics has the power to analyze serum antibody *diversity*, the breadth of antibody *binding* is most effectively analyzed using serological profiling by peptide and protein microarrays, which are devices with up to millions of peptides[233,234] or proteins[235,236] from either self-peptides, disease agents or random-sequence peptides.. Antibody binding to peptides is measured via fluorescently labelled secondary antibodies leading to signal intensities which are high for peptides which are bound by large numbers of antibodies and low when few antibodies bind. Antibody profiling studies may be roughly divided in two types[150]. (Type I) Epitope mapping studies: here, peptide signal intensities are interpreted as a reflection of the peptides' function (eliciting antigen and/or epitopes) in the studied disease. These approaches, therefore, mostly use autoantigen or disease-dedicated peptide libraries to detect and identify potential disease-antigens[237,238]. Type II peptide microarray approaches interpret measured signal intensities in relative terms and antibody polyspecificity[239]. The relative differences in binding patterns between the healthy and the diseased case represent the main finding. The fact that polyspecificity of antibodies renders the nature of the eliciting antigen(s) unimportant makes random-sequence peptide arrays, which are primarily used in this type of approach, a relatively cheap, unbiased, non-pathogen-restricted, and user-friendly tool for serological diagnostics[240,241]. Indeed, it was even hypothesized by Navalkar and colleagues that random-sequence peptides are more useful than tiling proteome sequences[242] for immunodiagnostic purposes. The reason for this may be that common motifs and protein structures are shared among many distantly related organisms. It may even be that in some cases, random sequence peptides have a higher number of 5-mer sequence overlap with the target proteome than peptide epitopes thought to be related to the target epitome[184,242].

Recently, a potential alternative to peptide arrays has been developed: PhIP-Seq (phage immunoprecipitation sequencing') combines oligonucleotide library synthesis with high-throughput DNA sequencing analysis of phage-displayed libraries. Compared with peptide microarrays, PhIP-Seq features longer and higher-quality peptides[243]. The synthetic oligonucleotide libraries are designed to encode peptide tiles that together span a library of protein sequences (entire proteomes, for example). The result is a representation of the encoded peptides. Deep DNA sequencing of phage-displayed peptidomes permits the quantification of each peptide's antibody-dependent enrichment thereby allowing the measurement of antibody specificity. PhIP-Seq has been applied to measuring the serum antibody against large viral peptide libraries[244] or to measure the repertoire of maternal anti-viral antibodies in human newborns[245].

One of the major remaining challenges of peptide arrays studies (or other serological profiling platforms) is the deconvolution of the binding signal exerted by each of the serum composing antibodies. Georgiev and colleagues showed that, at least in the context of HIV epitopes, the antibody binding signal of the serum circulating antibodies can be mathematically deconvolved to a certain extent if binding data exist of those monoclonal antibodies that are putatively in the serum. Current scalability issues of antibody expression (**Fig. 1**), however, render these experiments challenging on a larger scale.

While serological profiling represents a useful tool for mapping and uncovering antibody epitopes, mass spectrometry remains the only technology for comprehensively mining *in vivo* peptides presented to T cells via MHC. There has been considerable focus on leveraging mass spectrometry based imminopeptidomics to interrogate of HLA-peptide diversity and APCs for naturally processed peptides derived from tumor associated antigens[246]. Currently, there is an ongoing debate as to whether it is more effective to target hotspot drivers of cancer mutation or track private neo-antigen moving targets[247–249]. Identifying private neo-antigens has proven challenging, as identification requires the screening of enormous synthetic peptide, or fusion gene-libraries, and often involves procurement of rare and/or invasive patient samples. In spite of this challenge, several studies have developed personalized neoantigen vaccines which selectively expand antigen specific T cells and result in selected patient remission[250–253]. Of note, one of these studies was implemented in a phase I/Ib clinical trial for glioblastoma, a particularly low mutational burden tumor[253]. Together, these clinical success stories highlight the utility and impact of these immunopeptidomic approaches, and the potential to integrate deep learning to assist in the identification of (neo)antigens[254,255] and neoantigen-reactive T cells[256].

# Future directions for quantitative immunoengineering and immune receptor analysis

**Setting targets on public and private immune receptors**
Within the immune receptor landscape, instances occur where a given immune receptor is found in multiple people. These receptors are commonly called "public receptors" and they have been identified in health but also in numerous malignancies, infectious diseases, and autoimmune diseases[20,257–263]. Their detection requires brute force high-throughput paired-sequencing across large cohorts[8,9,56] or recently developed classifiers based on generation probabilities[54] and sequence motifs[91].

As of yet, the biological purpose of public clones remains unclear. It has been hypothesized that their function is to recognize and neutralize common pathogens[63,91]. However, definitive proof is still missing. We have previously published anecdotal evidence that murine B2-public clones show a higher overlap than B2-private clones with murine B1-B cells[91]. B1-B-cells are thought to bind common pathogens[264]. Relatedly, the Rajewski lab has showed that B1- and B2-B-cell lineage decisions depend on self-reactive signaling of the BCR. The inherent BCR-driven link between B1 and B2 suggests that the dynamics of the (public) repertoire potentially responsible for common pathogens is highly flexible and may not be entirely evolutionary encoded but rather environment-driven. In that regard, public clones may provide a pathway to identifying the functional limits of the adaptive immune system's target binding capacity as well as an indirect way of quantifying the frequency of antigen encounter throughout evolutionary time.

Apart from the fundamental immunological understanding of immune receptor publicity, its prediction may be important for population-wide targeting of immunotherapeutics[56]. Indeed, given the inherent complexity and overwhelming individuality of immune repertoires, public clones may represent predictable therapeutic "safe harbors" for antigen-specific targeting. Specifically, in recent years, vaccine design has begun to crack open the vision of targeting public clones for therapy by leveraging high throughput sequencing and structure guided design. In the case of HIV, high-throughput sequencing in a large-scale longitudinal study has identified BnAbs to HIV, which appear as public clonotypes, suggesting these public targets may be clinically valuable as they provide an opportunity to elicit by vaccination a few very potent therapeutic antibodies to treat many HIV infected individuals[258]. The capacity of public targets to independently arrive at the same clonotype against a given antigen, is an important demonstration of convergent evolution in immunology and suggests that public clones may act as crucial nodes in repertoire networks[77]. Germline targeting of condition-specific antibodies has also emerged as a promising competing approach to identifying antibody precursors with broadly neutralizing properties to modified immunogens[265,266]. Jardine and colleagues demonstrated using germline targeting through a highly interdisciplinary approach combining structure-guided protein interface design, yeast display libraries, and next-generation sequencing. Using this approach, they were ultimately able to develop a modified HIV immunogen with affinity for the germline-precursors of VRC01-class broadly neutralizing antibodies. While this germline targeting approach has been done in the context of antibodies, the work highlights an approach which could be co-opted to logically generate antigen specific T cells educated from on known private sequences and tailored immunogens. Germline targeting presents an opportunity to identify and expand condition specific TCRs and BnAbs where traditional methods fail to elicit these responses. In Focus Box 2, we discuss the possible implications the presence and absence of public BnAbs to HIV has on repertoire recognition holes.

Finally, in the context of T-cells specifically, public clones present themselves as an attractive target for engineered T-cell receptors and CAR-T therapy. Recent studies identifying public TCR recognition of the GAG peptide (protein encoded by retroviral genomes like HIV) revealed that these clones also display a high degree of promiscuity in their ability to bind many diverse HLA-DR allomorphs[257]. The ability of public TCRs to recognize an array of HLA, highlights the utility of public TCRs as this could allow some clones to be utilized across an array of patient HLA haplotypes. Related to that, the Friedman lab has identified, among peripheral T cells, MHC-independent public TCRβ CDR3 sequences that are abundant with fewer N insertions than average[267]. And very recently, the Brusko lab provided evidence that thymic selection is involved in the formation of public TCR clones[21].

**Efficient modification of immune receptor activity *in vitro* and *in vivo***
The humoral immune response elicited by vaccines efficiently triggers B-cell activation, somatic hypermutation, class switching, and the development of long-lived plasma and memory B cells which deliver enhanced protective and neutralizing responses[268]. It is also known that protective humoral immune responses are not always triggered by affinity maturation of the immunoglobulin repertoire, and that vaccines for common pathogens could not be developed. Even with identified sequences for protective monoclonal antibodies from many of these pathogenic targets, developmental vaccines often struggle to recapitulate the broadly neutralizing antibody responses present in elite controllers[269]. Using CRISPR/Cas9 mediated homology directed repair (HDR), it is now possible to introduce by targeted integration B-cell transgenes into hybridomas or primary human B-cell lines and exchange antibody specificity. Proof of principle of reprogramming the antigen specificity of B cells using modern genome-editing technologies was first demonstrated by the Reddy lab using the plug-and-(dis)play platform[270]. Here hybridomas were designed to remove endogenous copies of heavy and light chains, and full-length light and heavy chains, under the native IgH promoter, were expressed as a single transcript. Similar hybridoma CRISPR/HDR reprogramming has expanded upon this idea to generate on-demand class switching of recombinant monoclonal antibodies[271].

Recently, this concept has been applied to directly introduce well characterized protective paratopes of broadly neutralizing HIV antibodies via CRISPR/Cas9 mediated HDR into mature B cells. The goal of CRISPR-based lymphocyte engineering is to circumvent current challenges of generating HIV neutralizing responses via vaccination[269] or directed germline expansion[265,266]. The study is the first to describe a successful genome editing approach for introducing novel antibody paratopes into the human repertoire through BCR modification in polyclonal human B cells[272]. This approach replaces existing rearranged heavy chains with protective heavy chain paratopes derived from HIV broadly neutralizing antibodies. Recently two additional groups have independently developed alternative approaches utilizing CRISPR/Cas9 to reprogram murine and human B cells to also generate bnAbs to HIV[110,112]. While both groups demonstrated the utility of their approach using both naive primary human and mouse B cells, Taylor and colleagues demonstrate their system can reprogram B cells to generate bnAbs for not only HIV but also respiratory syncytial virus (RSV), influenza, and EBV. The B cell VDJ locus is also uniquely designed to retain physiological expression of the inserted monoclonal antibody and natural isotype class switching[112]. These preliminary results also suggest that after intranasal challenge with RSV, naïve B cells with genetically engineered receptors may be capable of establishing long-lived *in vivo* protection through generation of plasma-like and memory-like B cell populations. While initial results from these studies are promising, robust surface marker and transcriptional characterization of these cells is needed to verify that these cells are indeed long-lived, quiescent, and terminally differentiated B-cell populations[273].

Engineered TCRs have emerged as a promising immunotherapeutic strategy. Unlike antibodies, T cells lack the ability to undergo somatic hypermutation (with sharks being one of the few exceptions[274]), which contributes to a higher target specificity on the part of antibodies[128]. To better emulate the affinity maturation process, present in antibodies, many groups have attempted to develop ways which mimic this process for T cells. In recent years, there have been numerous studies to increase the affinity of a T cell to target antigen[275,276]. Unfortunately, clinical utilization of some engineered T cells has inadvertently shown that enhancing target affinity magnifies a latent capacity for TCR cross reactivity to non-target epitopes. For example, in one instance an affinity enhanced TCR was cross reactive to a peptide derived from the muscle protein Titin, causing lethal cardiovascular toxicity[277]. To combat this issue, it has recently been suggested that TCRs may be made to be more specific to a target antigen via the incorporation of both positive and negative design[278]. A recent review has expanded upon this idea to suggest that structural and predictive modeling will be essential in guiding design to identify these key mutations to maximize binding and specificity of TCRs[131]. Utilizing this approach would require positive design mutations with strengthening interactions between TCR and its cognate antigen/MHC, while introducing additional negative mutations which weaken evolutionarily conserved amino acids that control TCR-MHC interaction[279,280]. Indeed work from Brian Baker has led to ways to incorporate structure guided design to tune TCR affinity as well as enhance specificity independent of affinity[128,276]. In the latter study[128], positive and negative design was utilized, to our knowledge, for the first time to rationally modify TCR specificity and deliberately eliminate cross-recognition to non-target epitopes.

The interaction of TCR with peptide/MHC complexes, as well as CD4, CD8 co-receptors presents an additional opportunity to modify these complexes to enhance activity, independently of the TCR. Altering MHC anchor residues by generating heteroclitic peptide modification has been used as a common strategy to modulate immunogenicity *in vivo*[281–283]. Recent studies using CD4 yeast display libraries have identified highly conserved amino acid residues (glutamine at position 40 and threonine at position 45) which are essential for controlling interaction of CD4 with MHCII[284]. Mutation of these residues to tyrosine and tryptophan respectively, is capable of increasing responsiveness of low affinity TCRs to peptide-MHC (pMHC) complex. Importantly, these high affinity CD4 mutants appear to elicit increased reactivity to MHCII, resulting in stabilization of the pMHC-TCR complex to specifically increase responsiveness to cognate antigen and not non-specific antigen responses[285]. In light of this, it may also be possible to increase TCR activity by manipulating amino acids in the globular heads CD8αβ, or increasing binding avidity to MHC via glycan engineering of CD8 co-receptor stalks to generate high affinity variants; as high affinity CD8 coreceptors which lack these sugar residues are present naturally in double positive thymocytes[286,287]. While many of these techniques appear to be promising ways to enhance TCR sensitivity *in vitro*, they highlight the need to test ways to control activity in living organisms. Recent work from the Irvine group shows a biomaterial based delivery system that links IL-15 super-agonist laden nanocarriers to surface receptors of CD45, to selectively promote T cell expansion and tumor clearance by mouse T cells and human CAR-T cell therapy[288].

While the ability to heighten responsiveness of immune receptors is gaining traction, it remains equally important to maintain capacity to dampen or shut down engineered immune responses. Sophisticated drug controllable synthetic systems present themselves as an alluring, untapped means to modify immune receptor activity *in vivo*. Two competing methods relying upon drug controllable NS3 proteases have recently been described to reversibly control transcription factors, transmembrane signaling proteins, and drug stabilized variants of dCas9[289,290]. The ability to control protein activity with these hepatitis C virus protease inhibitors will undoubtedly have direct utility in the immunotherapy world. As engineered adaptive immune cell therapies gain more traction, it becomes increasingly important to begin developing ways to generate transcriptional control of engineered cellular therapeutics *in vivo*. Specifically, this means maximizing sensitivity to target ligands, improving therapeutic responsiveness, incorporating self-destructible kill switches, and control of master regulatory transcriptional factors in T and B cell lineage commitment. Benchmarks to efficiently control immune cells *in vivo* are further outlined in **Fig. 1**.

### *De novo* design of immune receptor sequences

Alongside the idea of developing predictive tools for identifying high-affinity immune receptors and epitopes, is the notion that biological evolution can be co-opted to accelerate the pace of identifying orphan receptors for desired cognate ligands. Directed evolution of epitopes has widely been explored as a way to accomplish this goal. Often this is performed by means of saturation mutagenesis and affinity maturation either *in vitro* or *in vivo*[291–295]. Typically applied through phage or yeast display libraries, these approaches have been limited in their ability to display largely diverse mammalian antibody libraries or, in the case of MHC, require random mutational trial and error in order to be displayed on the surface. One simple and versatile strategy employing site-directed mutagenesis utilizes the concept of CRISPR/Cas9-mediated homology repair to incorporate degenerate single-stranded oligonucleotides in murine variable heavy chain CDR3 to generate large libraries ($>10^5$) without the need for plasmid expression vectors, or PCR mutagenesis[270,296]. While libraries generated through homology-directed mutagenesis (HDM) are still several orders of magnitude lower than what can be routinely achieved with phage and yeast display systems, the system importantly demonstrates its ease of performing deep mutational scanning (DMS)[297,298]. In a similar approach to DMS, directed mutagenesis has also been directly applied to emulate an *in vitro* affinity maturation process in low affinity naive B cell repertoires[299]. DMS, HDM, or similar *in vitro* affinity maturation processes show the potential of providing meaningful insights to relate antibody function to sequence and may be capable of optimizing affinity and specificity of mammalian antibodies and TCRs. Although promising, it is important to recognize that these approaches should all be paired with cross reactivity screening to eliminate potentially autoreactive clones, similar to affinity maturation limitations imposed by central and peripheral tolerance.

To control for artificial positive and negative selection, methods leading to the generation of secondary lymphoid organoid complexes, which emulate artificial thymic and germinal center development, may be co-opted as powerful tools to direct evolution of immune epitopes. Artificial thymic organoid (ATO), described by Seet and colleagues, demonstrate a robust 3D system capable of generating mature naive antigen specific T cells through *in vitro* differentiation and positive selection of T cells[291]. T cells used in ATO system can also be generated from hematopoietic stem cells obtained from clinically relevant patient bone marrow biopsies, highlighting the potential for this system to be used alongside development of engineered T cell therapies. Similarly, immune engineered organoids for the generation of immunoglobulins have also been developed to further understanding of B cell physiology, malignancy development, and immunotherapeutic screening tools[300,301]. These artificial secondary follicles can be generated with tunable control over immunoglobulin class-switching but require starting with digested lymphoid tissue (rather than HSC) obtained from a patient's bone marrow. Artificial secondary lymphoid organoid systems have primarily shown their ability to emulate human T- and B-cell selection and maturation and play an important role in bridging developmental questions for *in vitro* and *in vivo* models. Future iterations of these immune organoids also open up the potential for more advanced systems to control, manipulate, and artificially focus human immune repertoires to accelerate the development of personalized immunotherapies.

Synthetic constructs such as bispecific antibodies[215], CAR-T cells and protein scaffold antibody mimetics (ex: DARPin, Avimer, AdNectin, Anticalin)[302,303] are major innovations in the development of *de novo* design of immunological variables. The synthetic design of these constructs has allowed them to compensate for natural failures of the immune system and have begun closing the gap between translating promising repertoire sequences, to clinically approved therapeutics. While development of engineered therapeutic systems like these are logical end states of de novo design of antibodies, their design, validation[304] and limitations is beyond the scope of this review. Moreover other reviews have already covered Antibody fragments[215], Bi-Specific antibodies[215], nanobodies[305], protein scaffold antibody mimetics[302,303], heterodimeric chimeric receptors (HCR T)[306–308], and CAR-T therapies[309,310], which we point the reader to.

Beyond immunological and molecular understanding of immune receptor binding, large-scale antigen-receptor-antigen data may enable computational and machine-learning guided design of epitope and paratope sequences[88,311–316]. Computational design of antibodies aims to create new antibodies with biological activity at a much higher rate than traditional methods for antibody discovery, such as animal immunization and large-scale library screening. Here, we briefly outline three advancement in antibody design, all with different purposes and effects.

The Fleishman lab has recently developed guidelines for the computational design of antibodies based on the design of naturally occurring antibodies[315,317]. They developed an algorithm that uses information on backbone conformations and sequence-conservation patterns observed in natural antibodies to design new antibody binders. Importantly, the designed antibodies were very different in sequence from natural ones but had appropriate affinity and stability. Furthermore, they found that two types of modeling constraints were required to make stable and expressible antibodies: conformation-specific sequence constraints as well as the use of large backbone fragments that included CDR 1 and 2 simplify sequence and conformation space.

The Ofran lab has recently succeeded in the computational design of epitope-specific antibodies. With an approach that does not rely on a solved structure of the target, they implemented a computational approach that integrates statistical analysis with multiple structural models as well as a random forest classifier to predict specific residue-residue contacts[313]. For an overview of recent approaches to computational antibody design, please refer to Ref[313] (Table 1).

Finally, computational antibody design is not only about target specificity and affinity. For therapeutic use, developability of antibodies is of equal importance. Current problems in developability are summarized by Raybould and colleagues and include high levels of hydrophobicity and patches of positive and negative charge[318,319]. They derived guideline values for five metrics thought to be important for antibody developability based on values seen in post-phase-I clinical-stage (CST) antibody therapeutics: (1) the total length of CDRs, (2,3) the extent and magnitude of surface hydrophobicity, (4) positive charge and negative charge in the CDRs, and (5) asymmetry in the net heavy- and light-chain surface charges[318]. The work by Raybould and colleagues builds on the work by Jain and colleagues who first analyzed the above mentioned CST antibodies in a comprehensive analysis of 12 different biophysical assays in common use for developability assessment[320].

Although there has been significant progress in biology-driven computational antibody design both with respect to target specificity and developability, it remains unclear how to reconcile optimization of specificity and developability since both entities are located in addition to the Fc region, to a major extent, in the CDRs of antibodies. More fundamental work is needed to understand the underlying biophysical foundations. Even more disconcerting, efforts are almost exclusively focused on antibody-sequence based analyses. However, as Kanyavuz and colleagues have recently pointed out: "structural restrictions of the canonical topology of the V region impose limits on the ability of antibodies to encompass the breadth of possible target epitopes. Incorporation of uncommon chemical groups (such as carbohydrates, sulfates, heme or metal ions), use of proteins or reconfiguration of the topology of the antigen-combining sites by conformational isomerism might all compensate for the structural hurdles imposed by different antigens."[321]. It remains entirely unexplored how to perform antibody design by accounting for all possible secondary chemical additions and modifications. It may be that with sufficient amounts of data, the high dimensionality of the problem is grasped by machine learning.

Recently, there has been exciting machine-learning driven progress in protein folding and chemistry that is also very useful for in-silico immune receptor design. Indeed, deep learning continues to push the boundaries of protein design, engineering and biomedicine as summarized recently by Ching and colleagues[322]. Of specific importance for protein design are recently reported deep learning guided directed evolution experiments that enabled the discovery of a functional green fluorescent protein with mutations in non-trivial positions[323] demonstrating that the deep learning model was able to supply the directed evolution experiments with previously unseen local maxima in the fitness landscape. A7D, Google's DeepMind deep learning based *de novo* protein structure predictor, is presently the top performer on CASP13 (Critical Assessment of Techniques for Protein Structure Prediction)[324,325]. Furthermore, chemistry is currently leading in terms of productive utilization of machine learning for drug discovery[326]. Segler and colleagues trained a deep neural network on all reactions ever published in organic chemistry and showed that the neural network can generate small molecule synthesis routes that are equivalent to reported literature routes[327]. More generally, recurrent neural networks may in principle be used for any text, sequence or graph[328–330] generating task thus leading, in the future, to not only the generation of single sequences but entire adaptive immune repertoires. After all, sequence generation is merely a special case of repertoire generation[331]. However, we wish to note here a fundamental problem for the large-scale validation of machine-learning based antibody design[332]: the low throughput of antibody generation[333] (and related to that: the issue of antibody standardization[334,335]). It remains currently unfeasible to produce 10,000s or 100,000s of different antibodies at large scale for verification of antigen specificity or developability (**Fig. 1**).

**Closing the data gap between immune receptor sequence and cognate epitope for immune receptor and epitope engineering**

One-way library screens have shown to be invaluable biotechnology tools for identifying and interrogating TCRs, antibodies and peptides[336–338]. These brute force search methods designed to identify rare matches for a given target are single-target-focused and therefore narrow in their breadth of search for identifying matching pairs (binding partners). This identification bottleneck poses a significant challenge to pharmaceutical target discovery as well as the creation of large-scale training datasets of known B- and T-cell receptor sequence pairs for machine learning approaches. Given this limitation, library-on-library screens (two-way screens) may prove to be a more desirable biotechnology platform for generating large-scale datasets of receptor-epitope pairs. Although no specific reports on the development of immune receptor library on two-way library screens has been reported, protein-protein interactions by yeast mating has recently appeared as a possible platform which could be exploited for these applications[339]. Using synthetic MATa and MATα haploid cells, a high-throughput method to interrogate protein-protein interactions was devised. Yeast mating was reprogrammed to link interaction strength with mating efficiency by substituting the MATa and cognate MATα sexual agglutinin subunits with barcoded binding peptides. In its current form, the approach presents itself as a proof of concept analysis tool capable of characterizing 7,000+ distinct protein-protein interactions. Barcoded synthetic agglutination gene cassettes present itself as a unique library-on-library selection tool. When combined with machine learning approaches, this system could generate training datasets capable of advanced receptor-based epitope/antigen prediction with single-epitope and single paratope resolution. Library on library selection methods would be the beginning of an entirely new era for the understanding of receptor-antigen pair formation. Once in place, we may then be able to detect high-affinity epitopes which may not be able to pass positive selection but produce functionally high affinity immune receptors. Importantly, we can then also finally begin to understand cross-reactivity and immunodominance[340] (extent to which one target attracts binders). With the advent of high-throughput library-on-library screening, new insight may enable the development of standard protocols for determining cross reactivity, and immunological functional outcomes may be predicted from paratope and epitope sequences alone.

Beyond library screens, Eyer and colleagues have recently published a microfluidics approach, called DropMap, to measure antigen-specific affinity and secretion rate of 100,000s of antibody secreting cells[341]. If, in the future, coupled with high-throughput BCR and transcriptome sequencing, the relationship between receptor sequence and target antigen may be screened at unprecedented scale[342].

In the effort to understand the statistical relationships of immune receptor-antigen pairs, several databases have been developed that have gathered such information: VDJDB[343], Abysis[344], IEDB[345], ABDB[90], McPAS-TCR[62]. These databases are continuously maintained and updated and therefore a valuable resource for future machine learning efforts. Superseding these comparatively smaller databases, are major efforts such as the iReceptor database[7] or the Observed Antibody Space[346], in part led by the AIRR Community (AIRR: adaptive immune receptor repertoires) to build large-scale databases that gather most sequencing datasets published in a preprocessed (AIRR-compliant) format that enables repaid repurposing of publicly available data for custom and new data analysis[347,348]. In fact, it was recently shown by Krawczyk and colleagues[349] that out of 242 post Phase I antibodies, 16 antibodies with sequence identity equal or higher to 95% of both heavy and light chains were also stored in publicly available immune receptor databases. While the databases containing immune receptor-antigen pairing are useful for prospective therapeutics discovery, entire repertoires will be needed for training machine learning approaches to immune (or disease) state prediction.

**Challenges in machine learning analysis on immune receptor repertoires**

Machine learning analysis on adaptive immune receptor data may be divided into two distinct problems: predicting antigen specificity given a specific receptor sequence and predicting disease states given a repertoire of immune receptor sequences.

The prediction of antigen specificity for a single receptor can usually be given in the form of a typical binary classification problem – whether or not a given receptor sequence is specific to an antigen/epitope of interest. There are several options for which receptor regions to consider: it may be the full V(D)J receptor sequence or only parts considered most critical for specificity (e.g., CDR3). Furthermore, it may be based on a single chain (typically $V_H$ or TCRβ) or with recent single cell-based technologies paired $V_H$-$V_L$/TCRβ-TCRα. Finally, as an alternative to framing this as a problem of classifying antigen specificity based on receptor sequence, one can set out to develop joint computational models of paired receptor-epitope data.

Using one-hot encoding, amino acids of the receptor sequence can be considered as distinct categorical values and represented as binary vectors. Another option is to translate the receptor sequence to a space of physico-chemical properties as performed by Thomas and colleagues[350]. As the input data is of variable length (typically CDR sequence length), it requires padding or alignment. The input data may also be represented by k-mers, which are frequency profiles of shorter subsequences (where k quantifies the nucleotide or amino acid length of the subsequence).

Intriguingly, the main challenge for machine learning analysis of immune receptors is the low dimensionality of the feature space, which is known to distinguish a very large number of antigens. Briefly, since antigen specificity is mainly determined by a ≈5–20 amino acid long $V_H$/TCRβ CDR3 sequence, the recognition and classification of billions of different antigens is based on decision boundaries within a feature space of merely ≈5–20 dimensions (or ~100–400 dimensions after one-hot encoding the categorical amino acid values). The low dimensionality implies the presence of strong high-order dependencies between amino acids (dimensions), which makes the learning task very difficult (without strong high-order dependencies, such a large number of distinct specificities could not be represented in the very limited number of dimensions). These strong dependencies reflect that the receptor in reality binds as a 3-dimensional structure, where variation at any given position may have implications for the entire structure and where residues that are distant along the sequence can be close in the folded 3D structure. Indeed, the majority of antibody epitopes, for example is thought to be conformational[144].

With these high-order dependencies as a premise, a fundamental outstanding challenge for machine learning on immune receptors is to determine the smoothness of antigen specificity in sequence space – in which ways sequences can change while retaining antigen specificity. This will have major implications for the choice of receptor sequence encoding because the main purpose of an encoding is to define a set of dimensions (features) that capture smoothness of the desired response variable (antigen specificity). The sequence signal associated with a given antigen specificity will likely be distributed among a set of separate subspaces of the sequence space, although present knowledge of this is very limited. A useful approach forward would be to benchmark[351] the performance of alternative models and methods on simulated data reflecting different hypothetical forms of sequence signals of antigen specificity.

In terms of smoothness and regularization assumptions, previous work on machine learning of receptor sequence specificity can mainly be placed in two categories:
1) Approaches assuming local smoothness in full-dimensional sequence space. Here, receptors having high general sequence similarity (e.g. according to Hamming or edit distance) in a receptor region such as the CDR3 are assumed to share antigen specificity, without any domain-specific consideration of the particular sequence discrepancies[24,83].
2) Approaches assuming that antigen specificity results from a combination of components in the form of receptor subsequences (contiguous or gapped k-mers), akin to natural language where semantics arise from a combination of words according to a given grammar[82,91,95,98]. In biological terms, one assumes the existence of receptor subsequences contributing defined binding characteristics that are not specific to a particular complete sequence context for a core region like CDR3. One could assume that components (subsequences) have defined binding characteristics independent or dependent of the position within a region like CDR3, which would imply encoding of a receptor sequence by k-mer frequencies or by k-mer-at-given-position frequencies, respectively.

Note that the use of k-mers does not in itself imply the second assumption. For instance, Glanville and colleagues[83] classify antigen specificity using in part a distance function based on the full k-mer frequency profiles of a pair of sequences. By incorporating the frequency of every k-mer, it is effectively a measure of the full sequence similarity, and thus according to the first assumption.

Machine learning is often contrasted to traditional statistics as being guided by general rather than specific priors. This means that instead of incorporating specific a priori expert knowledge about the domain, machine learning is implicitly guided by regularization priors that are shared by most prediction problems. According to such a perspective, different encodings, methods and architectures can be seen as reflecting different underlying regularization priors (smoothness assumptions). Comprehending antigen specificity smoothness in sequence space and developing well performing machine learning models is thus two facets of the same problem – with the first being a useful entry point from the direction of experimental immunological research and the second one being from the direction of benchmarking and machine learning research.

The prediction of disease states given a repertoire of receptor sequences is at its core a multi-instance, multiple-label learning problem. The term multi-instance denotes that the input data consists of a large set of receptor sequences, of which only an unknown subset is relevant for a particular disease. The term multi-label denotes that what is to be predicted is the presence/absence of multiple distinct disease states for each repertoire. Although one can always focus on a single disease and frame the problem as a binary classification of this disease state, a repertoire will inevitably contain receptors relevant for a variety of disease states, where there is no obvious way to filter out receptors not relevant for a single considered disease. Additionally, it is of importance that a learnt discriminator is representing the intended disease state rather than confounding factors (such as the age or the genetic or environmental background of the patients from whom the data were obtained). Both the multi-instance and multi-label characteristics contribute complexity to the prediction problem[352]. Furthermore, all challenges discussed for machine learning on individual receptors essentially carry over to the repertoire setting.

An outstanding challenge for prediction of disease states of repertoires is to integrate aspects of antigen specificity and V(D)J recombination probability. A fundamental assumption underlying machine learning approaches is that antigen specificity contributes selection pressure to a repertoire, allowing antigen-specific receptors to be detected through analysis of sequence or motif enrichment. However, rather than operating on a uniform sequence background, this selection pressure operates on the highly uneven sequence generation probabilities of V(D)J recombination. A fundamental challenge is thus to delineate sequence enrichment due to underlying generation probability versus antigen-based selective pressure. Such delineation would allow much more accurate estimates of the strength of selective pressure operating on a given sequence (motif). Furthermore, since the germline affects V(D)J recombination probability, the sequence generation probability could function as a mediator for confounding factors associated with a group of donors defined based on a trait of interest. Controlling for sequence generation probability could thus allow to break the connection to germline-associated factors, allowing both accurate and non-confounded estimates of antigen-based selective pressures.

While the discovery of antigen-associated patterns in receptor sequence is a relatively new field, the discovery of patterns in biological sequences in general has long roots more than forty years ago[353]. The discovery of sequence motifs representing DNA regulatory elements[354] has clear similarities to the discovery of receptor motifs, with the main difference being that regulatory element patterns are of lower complexity and typically embedded in longer sequences. It could be particularly relevant to draw inspiration from how the DNA motif field explored the difficulty of the problem and the performance of proposed methods. A very appealing approach is the one taken by Tompa and colleagues[355], where the authors of several competing methods came together to benchmark their methods on a collection of experimental datasets. However, due to the high variability of method performance across datasets and the limited total size of available data, the authors were not able to draw clear conclusions. Keich and colleagues[356] followed an alternative perspective, where they instead sought to inform on the problem from the theoretical side by delineating how subtle patterns could be before becoming indistinguishable from background. A hybrid perspective was employed by Sandve and colleagues[357], where a benchmark suite was created by assigning experimental datasets to three distinct groups according to an assessment of pattern complexity and Bayes error. The receptor bioinformatics field faces many of the same challenges, related to undetermined pattern characteristics, undetermined signal strength and limited availability of ground truth datasets. The field could thus get a head start by systematically building upon the experiences made in the field of regulatory element patterns.

Once a machine learning model has been decided on, it is crucial, especially in biomedical applications, to be able to interpret the results and the reasoning behind them[358]. In contrast to simple, linear models, the reasoning is not straightforward for more complex models or deep learning. Lanchantin and colleagues attempted to understand machine learning results by highlighting data which influences the decision the most[359] and Quang and Xie described a method that map the learned values to standard representations[360]. Furthermore, there exist software packages available which aim to extract the most significant features (see Focus Box 1). In the case of immune repertoires, the retrieval of such features would reveal the patterns characterizing the disease or antigen-binding and provide further insights for diagnostics and drug development. How to extract disease and antigen-specific motifs from immune receptor data remains largely unresolved.

In summary, given the complex nature of immune receptor repertoires, it may be of importance to integrate both sequence and frequency information as well as network similarity and phylogenetics frameworks[361,362] into future machine learning architectures in an effort to improve signal detection and decrease the influence of bystander noise.

**Relating immune receptor antigen specificity to cellular transcriptomic profile**
Currently, there exist very few studies that have linked immune receptor sequence and transcriptome to antigenic site specificity. Given that the transcriptome can be linked to cellular dynamics, linking the transcriptome with antigen specificity is important for understanding variation of antigen specificity on the cell subset level and cross-organ level. Specifically, it remains unknown how transcriptional programming correlates with epitope specificity across developmental stages, organ locations and disease states (health, cancer autoimmunity, infection). Indeed findings by a recent study on intratumoral T cells (breast cancer) suggests that T-cell phenotypes are likely shaped by a combination of antigenic TCR stimulation and environmental stimuli[363]. If transcriptome and antigen specificity are causally linked, then it may be conceivable to engineer higher affinity clones not by direct changing of the immune receptor but by changing key regulatory pathways, for example.

A crucial step in relating sequence to function is integrating transcriptomics with sequence chain pairing. The ability to subgroup and classify cells based on expression profile and paired antigen receptor information is an essential data factor for the mining of acumen and meaningful patterns in immune receptor sequences. While examples of integrating transcriptomics and chain pairing may be found in commercial systems such as that of the 10xGenomics platform, and other single-cell emulsion based assays, they are troubled by sequence read depth complications[364]. Lymphocytes appear to have low levels of mRNA transcripts of their heavy and light chains, and contain approximately 100 mRNA transcripts per adult B cell[365]. Furthermore, individual cells often lack consistency in sequencing depth[364,366], which is needed to accurately discriminate common markers (CD4, CD8, CD44, B220, CD62L etc.) to group receptors on a population level[367]. Development of high throughput screening assays capable of retaining chain-pairing and transcriptomic information will be a crucial next step to dissecting antigen receptor sequence and the functional outcomes of the cells. Such an assay could link disease specific effector cells to receptor epitope and would likely provide crucial insights into peptide specific motifs and predicting function from amino acid sequence.

Over the past few years, in addition to the 10xGenomics platform, there have been a few papers reporting computational tools that can extract immune receptor information from RNA-seq data[368–374] and applied to T-cell fate mapping in *Salmonella* infection[368] or relating clonal expansion to T-cell exhaustion state[375]. Taken together, linking transcriptome and immune receptor specificity is important to link molecular and cellular immune cell components for improved understanding of the microevolutionary processes that govern antigen-specific adaptive immunity.

## Conclusion

Since its inception nearly a decade ago[34,35], the quantitative analysis of immune receptor repertoires has enabled deep insight into the molecular basis of adaptive immunity. Many challenges remain to be addressed on each aspect of the investigation of immune repertoires to enable predictive immune receptor repertoire engineering and analysis: biotechnology, genomics, proteomics and computational immunology, machine learning and pattern mining. These challenges revolve around the one major knowledge gap of current research: the lack of insight into the immunological function and specificity of each immune receptor. Billions of immune receptor sequences stored in public databases are virtually useless for biomedical research if their specificity remains unknown. This knowledge gap pervades most current repertoire studies leaving behind, regardless of the study size, a hint of immunological inconclusiveness. Increasing our knowledge of the immune receptor's function will provide a gold standard and ground truth data for *in silico* diagnostics and therapeutics discovery. As both the public as well as funding agencies are invested in the design and analysis of real-world data, separation of noise from immunological signal will even more so rely on detailed understanding of the immune receptor-antigen interaction map, the immune grammar. Once resolved and amplified with next-generation computational immunology and machine learning, the possibilities for a multidimensional understanding and application of adaptive immunity will hopefully transcend our current imagination.

Finally, while we hope to have shown the great potential quantitative immunoengineering solutions harbor for the advancement of immunology, biomedicine and (digital) immunobiotechnology, we would like to caution against the ongoing separation of basic [experimental and computational] immunology and immunoengineering. In a recent opinion article for PNAS on "Why science needs philosophy", Thomas Pradeu and colleagues cited the late Carl Woese[376]: "a society that permits biology to become an engineering discipline, that allows science to slip into the role of changing the living world without trying to understand it, is a danger to itself."[377]. Indeed, only when merging fundamental immunology and immunoengineering, we may be able to understand, recreate, repair and improve adaptive immunity in a rule-driven fashion.

## Conflicts of Interest

E.M. declares holding shares in aiNET GmbH. V.G. declares advisory board position in aiNET GmbH. The remaining authors have no conflicts to declare.

# Focus Boxes

**Focus Box 1: Brief summary of deep learning and its architectures.**
A prominent class of machine learning algorithm is artificial neural networks (ANN)[405]. Typical ANNs comprise an input layer, a single or several hidden layer(s), and an output layer. Each layer comprises a set of nodes; nodes comprising hidden layers are termed hidden nodes. Theoretical studies on the approximative capabilities of these networks have shown that even a network with a single hidden layer may function as a universal approximator[406,407] (a function that maps any input to any output with sufficiently accurate approximation), albeit, with enough hidden nodes.

A neural net that has only a single hidden layer is known as 'shallow'. Early networks, for instance a multilayer perceptron or feedforward neural network[408], were setup in such a way that the set of nodes in the input layer is connected to the set of nodes in a single hidden layer. Finally, the hidden layer is connected to the output layer. Computationally, the 'connection' is made by a parameter *weight*. Intuitively, inputs are transformed (weighted) by the hidden layer to produce the outputs.

A more complex network comprising several hidden layers is referred to as 'deep' network. In biomedical applications as well as chemistry, deep neural networks have seen great success recently ranging from the prediction of transcription factor binding[409], skin cancer[410] and entire conception of chemical synthesis pathways[327]. Networks of this type may also include different types of hidden layers/nodes[411–413]. Recently, 'very deep' architectures (16 layers or more) demonstrated superior performance to their deep counterparts on a set of benchmarking image recognition datasets[414,415]. However, going deeper may not be the only solution in the quest for optimal performance[416,417].

Both shallow and (very) deep networks *learn* the weights from the input data, for deep networks this process is termed deep learning. Opportunities and challenges on applying deep learning in biology and medical research have been described in length elsewhere[322]. As of yet, it remains unclear as to exactly why deep learning is so successful in classification, prediction and generation. Ongoing research suggests that deep neural networks are especially adept at amplifying class-specific signal and ignoring class-unspecific noise[94], however, mechanistic details on what to amplify or ignore are obscured by the complexity of their architecture akin to a black box. Efforts to deconvolute deep and very deep networks (turning the black box into a glass box) have gained momentum in recent years, tools and techniques such as Integrated Gradients, LIME, and human-in-the-loop[418–420] are all geared towards attributing the prediction of a deep network back to its inputs.

**Focus Box 2: Recognition holes in the immune repertoire**
Holes in the immune repertoire were first hypothesized in 1971, and were initially thought to appear because of an inability of lymphocyte receptor gene mutations to generate target specificity[421]. As immunoengineering edges closer to de novo design of antibodies and TCR, it becomes increasingly important to understand where the natural boundaries of the immune repertoire lay, and their purpose. Incorporation of a model which predicts off limit V(D)J recombination could assist in developing engineered immune receptors with minimized cross reactivity. Concurrently, recognition holes may represent a beneficial target for immunoengineering by allowing the generation of rare pathogen neutralizing antibodies which are normally eliminated via central or peripheral tolerance.

Perelson and Oster asked how large a repertoire needs to be in order to be complete (no large holes in the repertoire). Specifically, they formulated the problem as follows: "given a set of *n* distinct, randomly made receptors, what is the probability that a randomly encountered antigen is recognized by at least one of the receptors?" To approach this problem, they introduced the concept of *shape space*[422]. Briefly, a receptor's shape is defined by the constellation of features important in determining binding to the antigenic space. A point in an *n*-dimensional space, ''shape space'' *S*, specifies the generalized shape of a receptor binding region. The shape space is a bounded region with volume *V*, assuming a restricted range of widths, lengths, charges, etc. that a receptor combining site can adopt. Complementarily, epitopes are also characterized by generalized shapes, which should lie within *V*. Because each receptor can recognize all antigenic determinants within a recognition ball, a finite number of antibodies may recognize a nearly infinite number of antigens thus giving a possible theoretical explanation of the broad recognition capacity. Specifically, Perelson and colleagues determined that a repertoire of $10^6$ randomly created receptors equates to 0.01% of escape epitopes. A caveat is however that these calculations were performed prior to the advent of high-throughput sequencing which has shown unequivocally that the immune repertoire is not created in a uniformly distributed fashion (let alone the predetermination imposed by the restricted germline gene repertoire)[3]. While the concept of the shape space has been instrumental in investigating repertoire holes, its main limitation is that it cannot be used for modeling of cross-reactivity, which is one of the main features of the adaptive immune system to increase its recognition breadth[239,423]. More recent mathematical approaches allow readily for the simulation of cross-reactivity[121,321,424,425] (see Section on *Mathematical modeling of immune receptor recognition*).

Central and peripheral tolerance appear to play a major role in the development of "recognition holes" across the development of the immune repertoire. Several studies have shown that a majority of immature human and mouse B cells generate autoreactive antibodies which are later eliminated through receptor editing[426,427]. Moreover, approximately 40% of antigen experienced IgG class switched antibodies from healthy human donors were also shown to have autoreactivity[428]. TCR repertoire depletion additionally occurs during thymic selection, whereby $CD8^+$ TCR repertoire size is inversely related to individual MHC diversity, causing an apparent tradeoff between MHC diversity and TCR repertoire size[101,429]. It is not known whether recognition holes exist entirely due to the presence of central and peripheral tolerance, or if specific amino acid combinations (sequence motifs) are capable of naturally evading immune monitoring. In TCR recognition, holes have appeared to be associated with human aging. Specifically, it appears that age associated decline in naive CD8 T cell diversity, may cause naturally low frequency precursors to become completely absent. This was shown to result in repertoire holes which compromise the ability to generate protective immunity[105,430]. HIV provides another interesting instance of where we believe holes in the repertoire exist alongside autoreactive antibodies. While broadly neutralizing human monoclonal antibodies (BnAbs) to HIV do exist, they are exceedingly rare. Many of these BnAbs have been found to have autoreactive properties *in vitro*[431]. In fact, it has become apparent that BnAbs to HIV often overlap with autoreactivity, and are disproportionately generated in patients with autoimmune diseases causing defective tolerance[432]. One report suggests that directed breaches in peripheral tolerance can naturally promote generation of these neutralizing HIV antibodies[433].

Moreover, approximately 8% of the genome is known to be composed of endogenous sequences with a high degree of similarity to infectious retroviruses[434]. Could it be that viruses partially exploit recognition holes generated by central and peripheral tolerance through becoming self over time? While this may be a potential consequence of tolerance, enormous holes in the recognition repertoire are actually more a feature rather than a bug of the immune system. Repertoire holes may provide essential checks on the development of autoimmune disease, and paradoxically can sometimes enhance protective antibody responses to foreign immunogens[435]. Given that recognition holes show correspondence to autoreactive epitopes which have overlap with pathogens, we wonder: Are there immunological settings that can selectively control tolerance to allow for the generation of BnAbs against specific pathogens? Future immunoengineering may seek to orchestrate breaches of self-tolerance in order to manipulate and expose potentially valuable "hidden epitopes".

**Table 1**: Summary of the most frequently used single-cell sequencing methods.

| Method name | Target | Single-cell technology | UMI use | Throughput* |
|---|---|---|---|---|
| Smart-seq[378] Smart-seq2[379] | Transcriptome | Micropipetting, FACS | No | Low |
| SMART-Seq v4/Fluidigm C1 | Transcriptome | Microfluidics | Yes | Low |
| MARS-seq[380] | Transcriptome | FACS | Yes | Low |
| Microwell-seq[381] | Transcriptome | Microwells | Yes | Medium |
| CytoSeq[382] | Transcriptome | Microwells | Yes | Medium |
| pairSEQ[383] | TCR | Microwells | Yes | High |
| Drop-seq[366] | Transcriptome | Microfluidics | Yes | High |
| 10x SC[384] | Transcriptome, TCR/BCR | Microfluidics | Yes | Medium |
| DeKosky (2013)[385] | BCR | Microwells | No | Medium |
| DeKosky (2015)[207] | BCR | Microfluidics | No | High |
| Briggs[209] | Transcriptome, TCR/BCR | Microfluidics | Yes | High |
| SPLiT-seq[210] | Transcriptome | Combinatorial barcoding | Yes | High |
| sci-RNA-seq[386] | Transcriptome | Combinatorial barcoding | Yes | Medium |
| Devulpally[387] | BCR | Manual pipetting | No | High |
| Busse[388] | BCR | Combinatorial barcoding | Yes | Medium |
| CEL-Seq[389] CEL-Seq2[390] | Transcriptome | Micropipetting Microfluidics | Yes Yes | Low Low |
| STRT-seq[391] STRT-seq2i[392] | Transcriptome Transcriptome | FACS FACS | No Yes | Low Low |
| scCAT-seq[393] | Transcriptome, epigenome | FACS | No Yes | Low |
| Zong[394] | Genome | Tissue lysis | No | Low |
| Xu[395] | Exome | Micropipetting | No | Low |
| scRRBS[396] Q-RRBS[397] | Methylome Methylome | Micropipetting Micropipetting | No Yes | Low Low |
| snmC-seq[398] snmC-seq2[399] | Methylome | FACS | No | Low |
| scATAC-seq[400] | Epigenome | Microfluidics | Yes | Low |
| scBS-Seq[401] | Methylome | FACS | No | Low |
| scWGBS[402] | Methylome | FACS | No | Low |
| scM&T-seq[403] | Methylome, Transcriptome | FACS | No | Low |
| scTrio-seq[404] | Methylome, Transcriptome, Genome | Micropipetting | No Yes No | Low |
| TetTCR-seq[224] | TCR | FACS | Yes | Low |

*Legend: Low: <10 000 cells per experimental run, Medium: 10 000–100 000 cells per experimental run, High: > 100 000 cells per experimental run

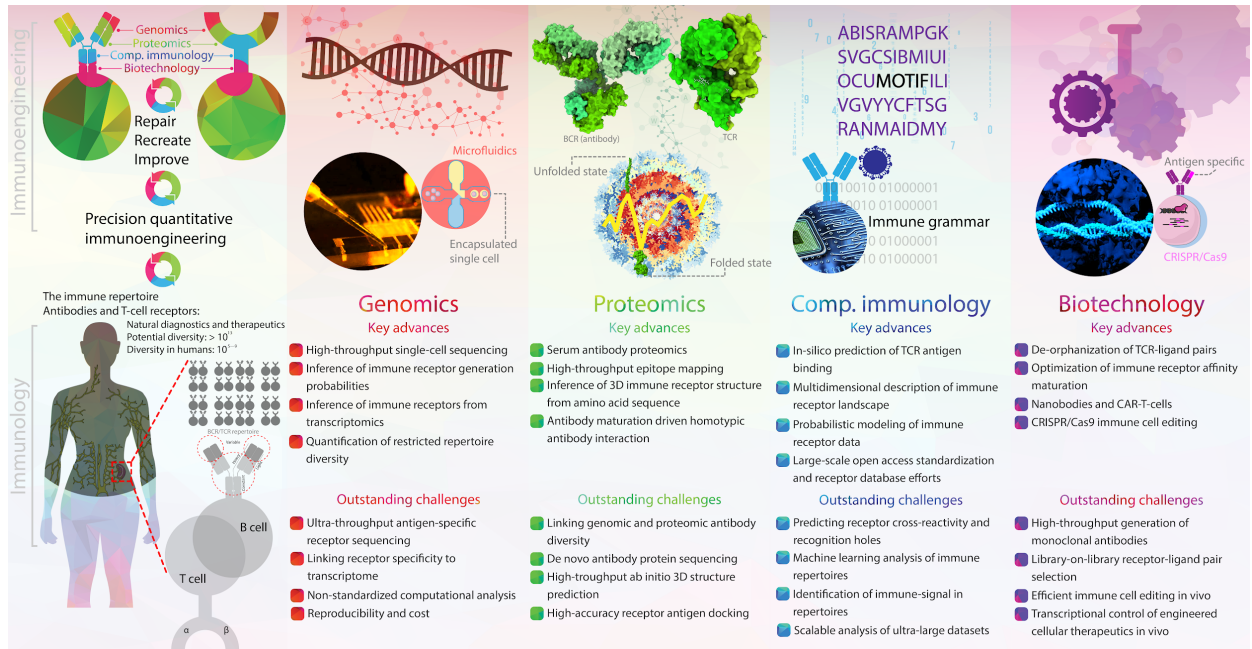

**Fig. 1: Key advances and outstanding challenges in quantitative engineering and analysis of adaptive immune receptor repertoires.** The immune system recognizes and neutralizes threats by evolving its sensors, B- and T-cell receptors, into a robust repertoire that is specific and at the same time broadly reactive. This is a highly potent system, but not without flaws. Pathogens sometimes slip through the cracks and an over- or under-reactive immune system manifests itself in the form of cancer or auto-inflammatory immune diseases. Recent advancements in both scale and sophistication of experimental and computational techniques have allowed experimental and digital immunoengineers to not only probe more complex questions, but importantly, it has ushered the era of augmented immunity. For the first time, the combination of ultra-large data and intelligent computational solutions opens a possibility to repair, recreate, and improve the immune system quantitatively and with great precision. Notable recent advancements such as the development of chimeric antigen receptor (CAR T cells) has indeed propelled the field forward considerably. However, a number of outstanding challenges remains to be overcome. Here we summarize the key advances in genomics, proteomics, computational immunology, and biotechnology along with some of the challenges. BCR and TCR molecular graphics were generated using UCSF ChimeraX[436].

## Notes and references